\documentclass[superscriptaddress,aps,prl,reprint,showpacs]{revtex4-1}
\usepackage[english]{babel}
\usepackage{graphicx}
\usepackage{amsmath}
\usepackage{amsthm}
\usepackage{amssymb}
\usepackage{soul}
\usepackage{float}
\usepackage{siunitx}
\begin{document}
\title{Time-resolved X-ray diffraction study of the structural dynamics in a ferroelectric thin film induced by sub-coercive fields
%Investigation of sub-coercive field induced structural dynamics in a ferroelectric capacitor via time-resolved X-ray diffraction
}
\author{C. Kwamen} \author{M. R\"ossle}
\affiliation{Helmholtz Zentrum Berlin, Albert-Einstein-Str. 15, 12489
  Berlin, Germany}
\author{W. Leitenberger}%\author{F. Zamponi}
\affiliation{Institut f\"ur Physik \& Astronomie,
  Universit\"at Potsdam, Karl-Liebknecht-Str. 24-25, 14476 Potsdam,
  Germany}
\author{M. Alexe}
\affiliation{Department of Physics,
  University of Warwick, Coventry CV4 7AL}
\author{M. Bargheer} \email{bargheer@uni-potsdam.de}
\homepage{http://www.udkm.physik.uni-potsdam.de} \affiliation{Institut
  f\"ur Physik \& Astronomie, Universit\"at Potsdam,
  Karl-Liebknecht-Str. 24-25, 14476 Potsdam, Germany}
\affiliation{Helmholtz Zentrum Berlin f\"ur Materialien und Energie, Albert-Einstein-Str. 15, 12489
  Berlin, Germany}

  \newcommand{\superscript}[1]{\ensuremath{^{\textrm{#1}}}}
\newcommand{\subscript}[1]{\ensuremath{_{\textrm{#1}}}}

\date{\today}
\begin{abstract}
The electric field-dependence of structural dynamics in a tetragonal ferroelectric lead zirconate titanate thin film is investigated under sub-coercive and above-coercive fields using time-resolved X-ray diffraction. During the application of an external field to the pre-poled thin film capacitor, structural signatures of domain nucleation and growth include broadening of the in-plane peak width of a Bragg reflection concomitant with a decrease of the peak intensity. This disordered domain state is remanent and can be erased with an appropriate voltage pulse sequence.
\end{abstract}

\maketitle
Ferroelectrics (FE) are not only technologically interesting because of their electromechanical properties that enable their application in transducer devices. The reversible spontaneous polarization of FE has been used in memory devices with a reported data retention of about 10 years~\cite{Scott1989}. FE memory devices make use of the remnant polarization state obtained after poling or reversing the FE polarization by an electric field to store the boolean algebraic logic states ``0`` and ``1``. Operating FEs above their coercive field leads to fatigue, which limits device lifetime~\cite{Duiker1990,Wang2002,Luo2012,Shieh2006,Jiang1993}. Also, the quest for low power consumption requires operating such devices under the lowest possible bias. Hence there is growing interest in sub-coercive field applications and associated remnant states~\cite{Kundys2014,Yimnirun2008,Pramanick2011,Jones2012,Wongdamnern2010}.\\
It is well known that FE thin films do not reverse their polarity as a whole when an external field is applied.
After domain nucleation, the regions with opposite polarization are separated by domain walls (DW)~\cite{Merz1954a, Kim2007}. One of the domain polarization directions prevails as the external field approaches the saturation field, however, the time required to fully suppress domains with the opposite polarization is determined by the domain wall velocity.  The DWs are mobile, add structural disorder within the sample volume, and reduce the observed maximum polarization. DWs contribute to the observed piezoelectric strain in FE devices already at very low fields~\cite{Pramanick2009a,Pramanick2009b}. This has lead to new devices in magnetic and FE materials for DW logic and DW diode applications~\cite{Allwood2004, Allwood2005,Whyte2015}. It has recently been demonstrated that multiple memory states can be created due to the coupling between remnant strain and domain states~\cite{Kundys2014,Zhou2016}.\\
Some actuator applications require fast actuation with reproducible length changes that are not affected by drifts on longer time-scale. To study the time-dependent response of FE ceramics, time-resolved X-ray diffraction (TR-XRD) has been used to probe the dynamics under electrical loading, offering the possibility to distinguish between the intrinsic piezoelectric response of individual domains and extrinsic contributions originating from changes of the volume fraction of the domains.~\cite{Pramanick2009a,Pramanick2009b}. In the last years, \emph{in-situ} synchrotron XRD has been used to quantify the electromechanical response and fatigue behavior of FE thin films and powders under loading, with the advantage of yielding appropriate temporal and spatial resolution as compared to conventional X-ray setups~\cite{Davydok2016,Do2008,Gorfman2015,Cornelius2017}. \emph{In-situ} high-resolution XRD has for example been used to study the switching dynamics of $90^{\circ}$ domains in epitaxial PZT~\cite{Jones2006}. More recent measurements aim at combining laser excitation and electrical excitation for x-ray structural dynamics investigations \cite{akam2018}. Several synchrotron-based experiments reported intensity and position changes of the diffraction maxima in PZT thin films, however, electrical characterization was studied independently \cite{grig2006PRL,grig2009PRB}. The interpretation of very large intensity changes up to $40\%$ for the two polarization states solely ascribed to the anomalous scattering must be reconsidered since contributions from disorder cannot be neglected. We have recently reported the structural dynamics accompanying the reproducible repeated electrical switching of FE thin films, relating the observations to the simultaneously measured electrical charging with saturation fields of opposite polarity~\cite{Kwamen2017}.

\begin{figure}%[H]
 % \centering
  \includegraphics[width=7.5cm]{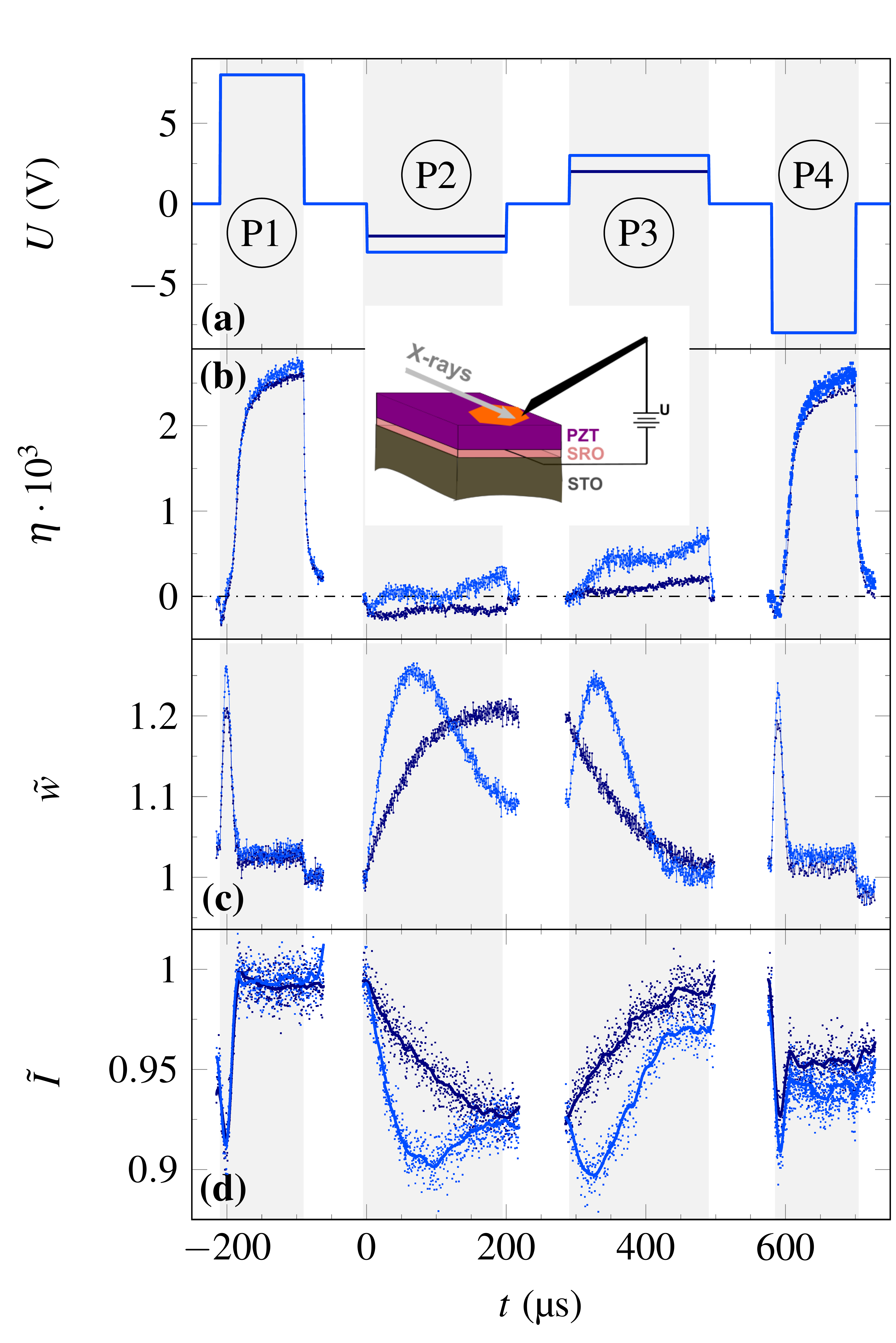}
  \caption{Time-dependence of $c$-axis strain $\eta$, normalized peak width, $\tilde w$, and normalized integrated intensity, $\tilde I$, during sub-coercive field loading. (a) Applied waveform, (b) out-of-plane lattice response to the applied field, (c) angular broadening of the measured peak, and (d) corresponding integrated intensity. The inset schematically shows the sample structure, the application of the voltage,  and scattering geometry.}
\label{fig:weak field}
\end{figure}

In the present TR-XRD study of a 250\,nm thin $\text{Pb}(\text{Zr}_{0.2}\text{Ti}_{0.8})\text{O}_3$ (PZT) FE capacitor, we compare the time-dependent structural dynamics during sub-coercive field pulses to the full switching dynamics under saturation fields. We present the field- and time-dependent lattice response, infer the domain dynamics, and demonstrate the presence of remnant disordered states.
%\subsubsection{Methods}
The experiments were realized on a FE capacitor composed of PZT as FE layer and a metallic SrRuO$_3$ (SRO) electrode epitaxially grown on a (001) oriented SrTiO$_3$ (STO) substrate by pulsed laser deposition.  Platinum top electrodes of hexagonal shape with 0.3\, mm edge length and a thickness of about 30\,nm were deposited onto the PZT film by sputtering. The sample structure is sketched in the inset of Figure~\ref{fig:weak field}. An external field was applied between one Pt electrode and the SRO layer along the [001] direction, which is the FE polarization axis of PZT. A pulse generator (Keithley 3390) was used to apply voltage pulses up to $\pm 8$ V with the help of a tungsten needle~\cite{Kwamen2017}. X-rays of 9\,keV photon energy provided at the XPP-KMC3 endstation at the storage ring BESSY II at the Hemholtz Zentrum in Berlin were employed for the TR-XRD. The experiments were performed in the so-called multibunch operation mode \cite{Holldack2014,Navirian2012}. The X-ray beam was focused onto the sample surface such that only the region under one electrode was probed. For recording full reciprocal space maps (RSM) a PILATUS 100K pixel detector (DECTRIS) was used. To obtain time-resolved signals, the detector gate was triggered appropriately by the electric field pulse sequence. In order to increase the acquisition speed, $\omega$ scans through the RSM were recorded with a home-built detector made of a photomultiplier tube (Hamamatsu) and a fast scintillator with \textless $1$ ns response time, which collects an angular divergence of the diffracted x-rays of $0.9^\circ$, which exceeds the width of the RSM in either angular direction by a factor of three. This detection geometry is often referred to as rocking curve, an established method to determine lattice plane spacings in mosaic crystals via Bragg's law~\cite{Ayers1993,Serafinczuk2016,Kobayashi1999}. The typical full-width-at-half-maximum (FWHM) of the rocking curves was $0.3^\circ$.%,Serafinczuk2016
The X-ray signal is sent to a time-correlated single photon counting system (PicoHarp 300), which allows reconstructing the rocking curves at different delays by synchronizing the detector trigger with the rising edge of the electrical pulse.~\cite{Navirian2012}
The high resolution X-ray diffraction measurement showed that the film is $c$-axis oriented~\cite{Kwamen2017}. The reciprocal space map without applied external field was reported in a recent publication for a different electrode of the same sample~\cite{Kwamen2017}.
%\subsubsection{Results and Discussion}
The excellent FE properties of the device with an $RC$ time constant of about 1\,$\mu$s and a coercivity of about 4.5\,V have been discussed in ref.~\cite{Kwamen2017}: Polarization reversal observed by electronic measurements was shown to be accompanied by a intensity changes of about $\Delta I = 5 \%$ resulting form difference in structure factor between the (00$2$) and (00$\bar{2}$) Bragg reflections ~\cite{Kwamen2017}. This is consistent with a relative motion of the Ti and Zr atoms relative to the Oxygen octahedra of $\Delta \xi_{\text{Pb-Ti}}=0.0187$\,nm and $\Delta \xi_{\text{Ti-O}}=0.03$\,nm along the $c$-axis~\cite{Noheda2000}. Now we apply an asymmetric square pulse sequence as shown in Figure~\ref{fig:weak field}(a) to a test device and analyze the dynamics of the 002 PZT reflection. The strain $\eta= \Delta c(t)/c(t=0)$ corresponding to the relative $c$-axis change is shown in Figure~\ref{fig:weak field}(b). In standard models of polarization switching in thin films under saturation fields~ \cite{Merz1954a,Kim2007,Dawber2005,So2005,Fatuzzo1959}, the polarization reversal is mediated by the nucleation of new domains with reversed polarization, their subsequent propagation across the film thickness, and a subsequent lateral motion of DWs perpendicular to the field direction that finally leads to a merging of the reversed domains. The structural dynamics observed in our sample for two different voltages, $U$, are shown in Figure~\ref{fig:weak field}. The pulses marked P1 and P4 are the poling pulses used to initialize the film\textsc{\char13}s polarization state in a reproducible manner. Pulse P2 and P3 are chosen to be symmetrical and below the coercive field $U_{\text c}$. Note that the time $t=0$ is defined as onset of pulse P2, since we are interested in the sub-coercive field dynamics.\\
We first discuss the darkblue lines in Figure~\ref{fig:weak field} representing an electric field of -2\,V, well below $U_{\text c}$ of the device. The PE response to pulse P2 is purely compressive (Figure~\ref{fig:weak field}b), indicating that the majority of the thin film has retained its initial polarization. The relative peak width $\tilde{w}=(w(t)-w(t=0))/w(t=0)$ represented in Figure~\ref{fig:weak field}(c) increases slowly and reaches a saturation value after $\sim\!150$\,$\mu$s with an overall width change of $\sim\!20$\%. We also note that the value of $\tilde{w}$ remains constant even after switching the external field off at about 200\,$\mu$s. Simultaneously, the normalized integrated intensity $I$ shown in Figure~\ref{fig:weak field}(d) decreases by $\sim\!7.5$\% of its initial value. The continuous change of $\tilde{w}$ at constant field shows that domains nucleate and move within the FE film towards a state with approximately equal amounts of both polarization states as this increases the inhomogeneity of the film. Pulse P3 with opposite polarity is applied after a short field-free time and recovers $\tilde{w}$ and $I$, thereby resetting the domain state in the film. The stability of $\tilde{w}$ during the time between the electrical pulses, i.e. at zero fields, proves that the disordered domain configuration induced by sub-coercive fields is remnant. In contrast, the small PE $c$-axis contraction, which is induced by pulse P2, relaxes back to its zero-field value. Pulse P3 with opposite polarity then leads to an expansion with similar amplitude, which is consistent with the small applied field below $U_{\text c}$, which leaves the majority of the FE film in the initial polarization state.
 \begin{figure}%[h]
  \includegraphics[width=8.7cm]{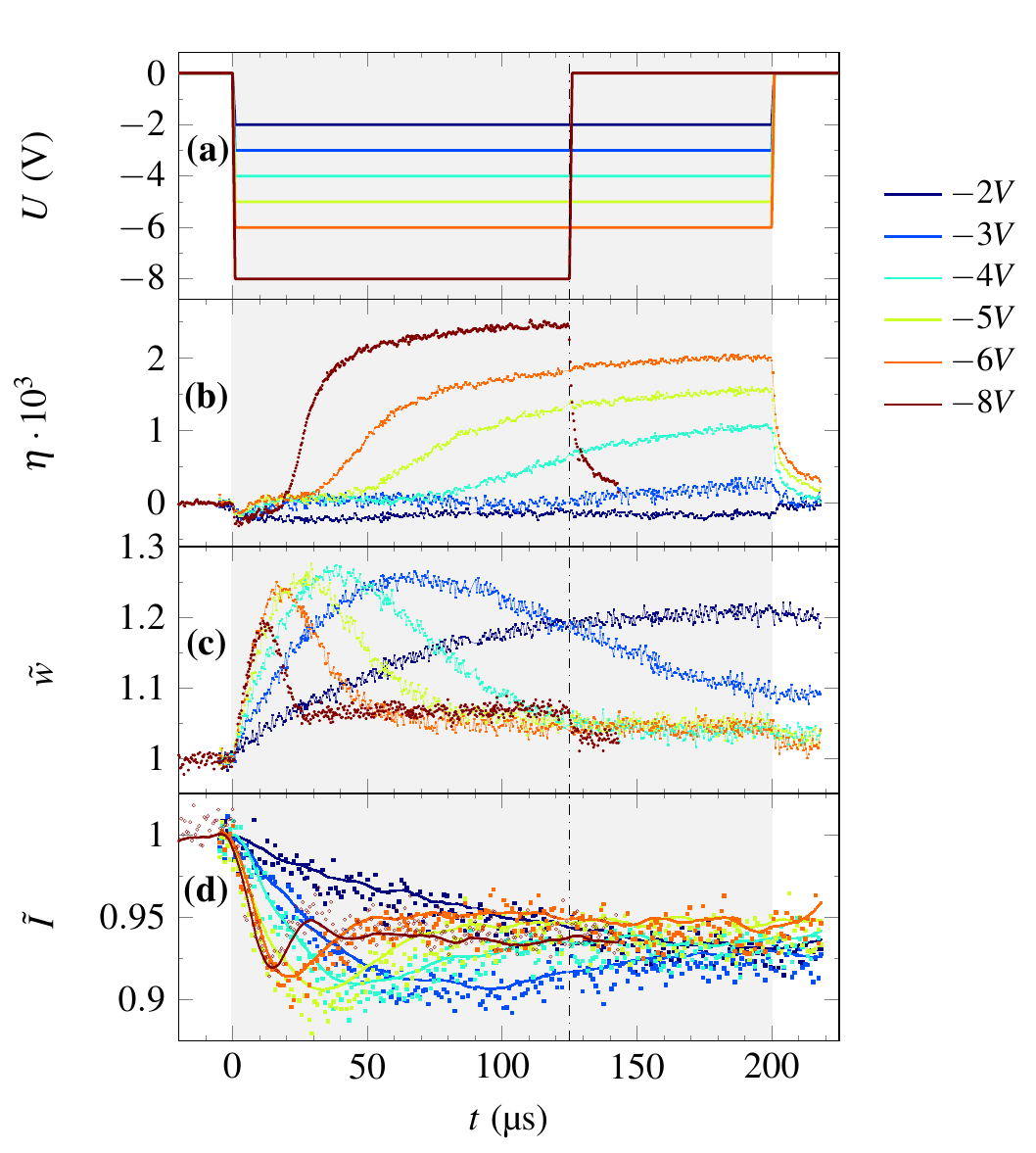}%
  \caption{Time-dependence of the $c$-axis strain, $\eta$, the normalized peak width, $\tilde w$, and normalized integrated intensity $\tilde I$ under switching pulses of varying amplitude $U$. (a) Schematic of the applied waveform, (b) strain due to the applied field as a function of time, (c) normalized  FWHM from the recorded rocking curves, and (d) corresponding integrated intensity normalized to its value at $t=0$. }
\label{fig:varying amplitude}
\end{figure}
We now discuss the blue lines in Figure~\ref{fig:weak field} that represent a somewhat larger electric field pulse with $U_2=-3$V, which is just below $U_{\text c}$ for a standard hysteresis loop measured at 1\,kHz. Now a PE expansion is observed during both pulse P2 and P3, indicating that a bit more than 50\% of the thin film have reversed polarization during pulse P2, which is then switched back by pulse P3. Thus for this pulse sequence, the coercive field is slightly below $3$\,V.\\
In Figure~\ref{fig:varying amplitude} we concentrate on the temporal sample response during Pulse 2 with various field amplitudes. In Figure~\ref{fig:varying amplitude}(a), the different applied fields are plotted as function of time after the application of the pulse. Before each of these pulses were applied, the poling pulse P1 from Figure~\ref{fig:weak field}(a) with opposite polarity and amplitude 8\,V set the initial condition of the sample. We varied the amplitude of pulse P2 from $-2$ to $-6$\,V and kept its duration constant. For the experiment at $-8$\,V we reduced the pulse length to avoid degradation of the sample. The X-ray analysis of the structural dynamics of the PZT film is plotted in Figures ~\ref{fig:varying amplitude}(b-d). Figure~\ref{fig:varying amplitude}(b) confirms the PE contraction of the film for $-2$\,V which can now be compared to the much larger inverse piezo-effect obtained under larger fields. In addition, we would like to highlight the steeper and steeper gradient of the expansion dynamics for larger fields. The domain nucleation and propagation is more efficient and leads to a more rapid positive PE expansion. The initial increase of $\tilde{w}$ of about 20\% shown in Figure~\ref{fig:weak field}(c) indicates disorder in the film that increases by domain nucleation and propagation. For fields $|U_2|\ge 3$\,V $\tilde{w}$ decreases again as domains merge and grow, thus decreasing the disorder. For $U_2=-3$\,V some disorder remains when pulse P2 comes to an end and this disorder remains under zero field for $t > 200 \mu$s. For $U_2=-2$\,V the pulse ends in the state of maximum peak width, and also this signature of large domain disorder is remnant. In fact, a small increase of the peak width remains after switching off Pulse P2 even for the largest applied field. This probably indicates a slight asymmetry of the film's structural properties. For the high field of $-6$\,V and $-8$\,V the normalized diffracted intensity plotted in Figure ~\ref{fig:varying amplitude}(d) confirms the expected structure factor change for full polarization reversal. However, for low field switching, the intensity decreases by more than the $\Delta I = 5\%$ expected for the structure factor change. The polarization is only reversed in a small fraction of the sample, and hence it is not the structure factor but rather disorder associated with inhomogeneous domain nucleation and motion that induces the intensity reduction.
\begin{figure}%[H]
  \centering
  \includegraphics[width=8.30cm]{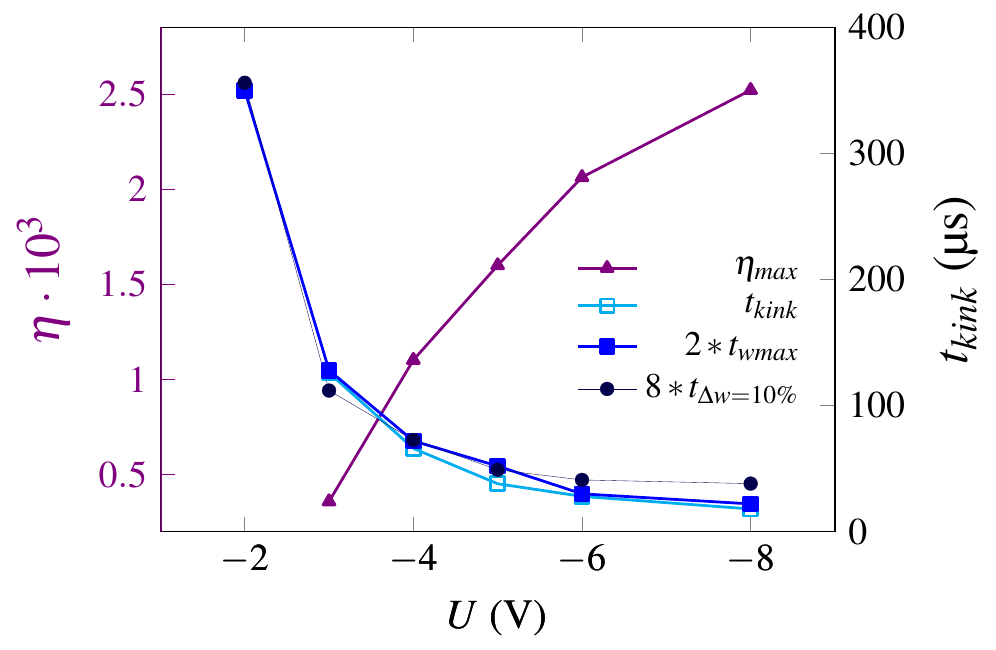}
  \caption{Maximum strain $\eta_{\text{max}}$ at $t = 200$\,$\mu$s as a measure of the FE polarization as function of the applied voltage $U$. The entire dynamics speed up with increasing voltage: We compare the times $t_{w10\%}$ it takes to initially increase the peak width by 10$\%$ and the $t_{w\text{max}}$ (blue solid squares) characterizing the time for the maximum peak width to $t_{\text{kink}}$ (open squares) at which the strain $\eta$ suddenly increases (filled black symbols). The times have been scaled by appropriate parameters given in the legend. }
\label{fig:meta}
\end{figure}
Now we extract the significant signatures of sub-coercive field induced polarization reversal from Figure~\ref{fig:varying amplitude}(b-d) and plot these signatures as a function of the field $U$ in Figure~\ref{fig:meta}. Each of the strain transients above $|U_2|\ge3$\,V in  Figure~\ref{fig:varying amplitude}(b) shows a pronounced kink where the rate of expansion, $d\eta / dt$, changes. It occurs at the time $t_{\text{kink}}$ which shows a very similar field dependence as the time $t_{\text{$w$max}}$ at which the maximum of the peak width is reached. This indicates maximum structural disorder %$t_{\text{kink}}$ and $t_{\text{max}}$ are represented in Figure~\ref{fig:meta} by the open and closed squares, respectively.
%  (see light and dark blue data in Figure~\ref{fig:meta}.
%We interpret $t_{\text{max}}$ as the time where the polarization reversal speeds up because
and the point where the reversed domains start merging. This time precedes the time $t_{\text{kink}}$ of the sudden increase of the lattice constant by a constant factor of $t_{\text{kink}}/t_{\text{$w$max}}=2$.
Figure~\ref{fig:meta} also reports the initial rate $dw/dt$ at which the peak width changes. In order to emphasize the identical voltage dependence we plot the time  $t_{\Delta w=10\%}$ where the peak width has increased by $10\%$ (full black circles) as a measure of the inverse rate. There is another constant factor of $t_{\text{kink}}/t_{\text{w10\%}}=8$.\\
We would like to emphasize that the different remnant values of $\tilde{w}$ after 200\,$\mu$s define domain states that are stable as long as no electric fields are applied. This disorder could be used for multi-level memory applications based on remnant domain configurations~\cite{Kundys2014}. Such domain states can be created by the pulse sequence discussed above.
\begin{figure}%[H]
  \centering
  \includegraphics[width=8.7cm]{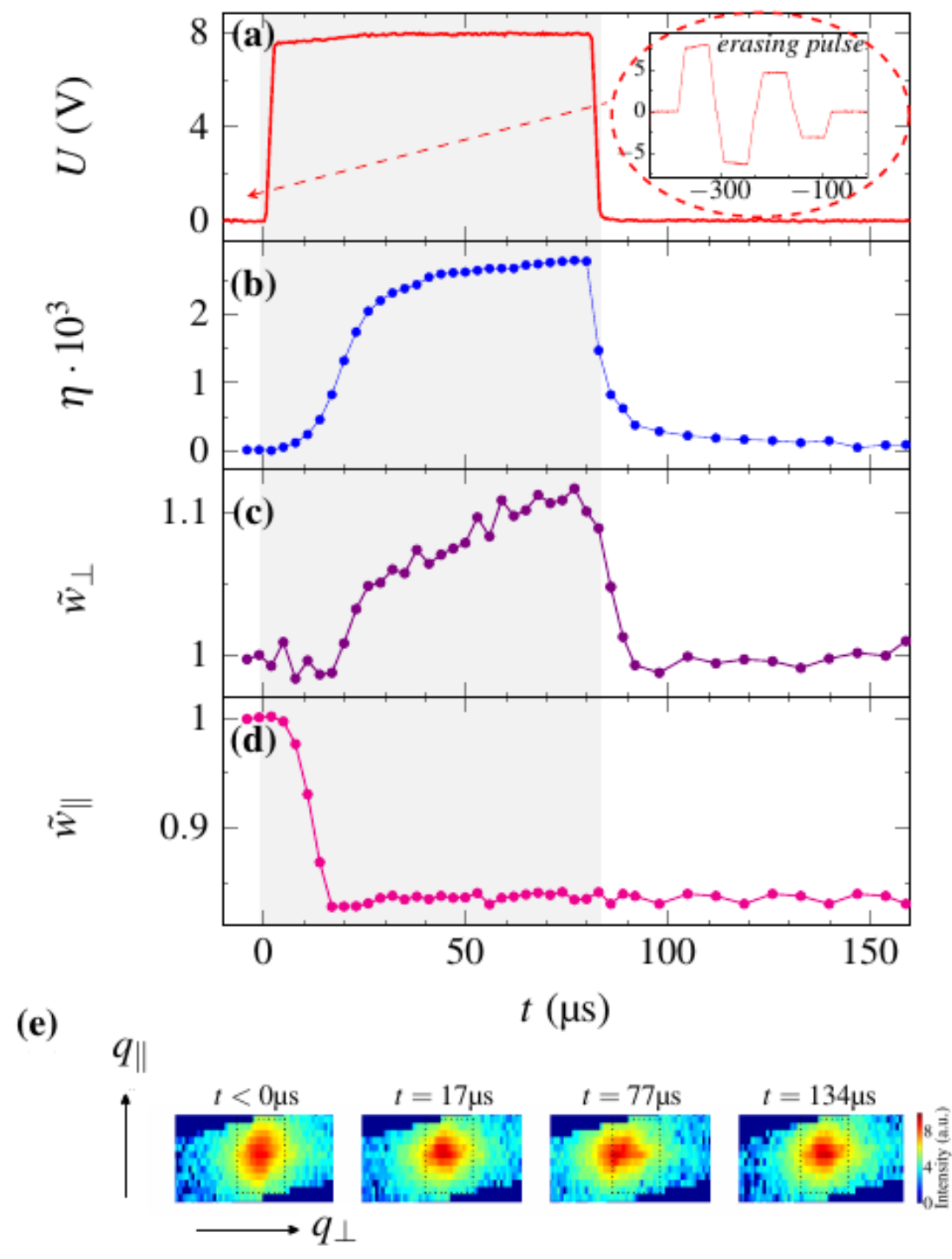}
  \caption{Structural response to an above-coercive field pulse directly after the ``erasing`` pulse sequence shown in the insert of panel a). (a) Applied pulse, (b) strain obtained form the change of the out-of-plane lattice constant. (c) Normalized FWHM of the out-of-plane Bragg peak. (d) Normalized FWHM of the corresponding scattering vector parallel to film surface. (e) RSMs at selected times. The width and the height of the displayed RSM corresponds to $\tilde{w}_{\perp}=0.09 \si{\per\angstrom}\ $  and  $ \tilde{w}_{\parallel}=0.07 \si{\per\angstrom}\ $ respectively.}
\label{fig:ErasePulse}
\end{figure}
In the following, we demonstrate that these states can be erased by using an appropriate pulse sequence in order to define a reference state for multi-level memory applications. A simple "erasing" pulse sequence is indicated in the inset of Figure~\ref{fig:ErasePulse}(a). We monitored this reference domain structure and the changes induced by the writing pulse shown in Figure ~\ref{fig:ErasePulse}(a) with 8\,V amplitude by reciprocal space mapping (RSM). Figure ~\ref{fig:ErasePulse}(e) visualizes the structural changes directly in the RSM for selected time delays. Figure~\ref{fig:ErasePulse}(b) shows the peak shift, which is only present while the electric field is applied. Figure~\ref{fig:ErasePulse}(c) indicates the normalized width $\tilde{w_\perp}$ along $q_z$ indicating an inhomogeneous electric field distribution across the electrode area.\cite{Kwamen2017} This peak width change disappears after the external field is switched off. In contrast, the large FWHM of the in-plane component of the Bragg peak $\tilde{w}_{\parallel}$ after the  ``erasing`` sequence indicates a high density of small domains, which is rapidly reduced by the external field pulse. This ordering of the domain structure again is remnant and states of different degrees of disorder are stable states in the thin film device.
%\subsubsection{Conclusion}

We demonstrated that states of remnant ferroelectric domain disorder can be written in a nanometric PZT thin film device at sub-coercive fields. The structural changes of the domain states were observed by time-resolved reciprocal space mapping with hard X-rays, looking at a working device that was operated for more than $10^8$ switching cycles during these measurements.  This sub-coercive field switching could be used in low power ferroelectric memory applications operating at low field. We showed that the intensity of Bragg reflections changes during the switching not only because the structure factor changes but also because the domain disorder broadens the Bragg peaks. \\%This is very important for the characterization of ferroelectric switching by combining electrical and structural methods~\cite{Kwamen2017}.
%\subsubsection{Acknowledgements}
We acknowledge the BMBF for the financial support via grant No. 05K16IPA.
 M.A. acknowledges support of Royal Society through the Wolfson Research Merit and Theo Murphy Blue-sky Awards and EPSRC (UK) through grants No. EP/P031544/1 and EP/P025803/1.

%\bibliographystyle{apsrev}
%\bibliography{references}
\bibliographystyle{apsrev}

%\bibliography{referencesSubEcN}

\end{document}